\documentclass[aps,prd,twocolumn,showpacs,preprintnumbers,superscriptaddress,ams
math,amssymb]{revtex4}
\usepackage{graphicx}
\usepackage{dcolumn}
\usepackage{color}
\usepackage{epsfig} 
\graphicspath{{ps}}

\newcommand{\gev} {\ensuremath{\, {\mathrm{GeV}}     }}
\newcommand{\mev} {\ensuremath{\, {\mathrm{MeV}}     }}
\newcommand{\gevc}{\ensuremath{\, {\mathrm{GeV}/c^2} }}
\newcommand{\mevc}{\ensuremath{\, {\mathrm{MeV}/c^2} }}

\newcommand{\ecm} {\ensuremath{ E_{\mathrm{c.m.}} }}
\newcommand{\gisr}{\ensuremath{ \gamma_{\mathrm{ISR}} }}

\newcommand{\RMS} {\ensuremath{ M^2_{\mathrm{recoil}} }}

\newcommand{\ee}     {\ensuremath{ e^+e^- }}
\newcommand{\dsp}    {\ensuremath{ D_s^+ }}
\newcommand{\dsm}    {\ensuremath{ D_s^- }}
\newcommand{\dsstp}  {\ensuremath{ D_s^{*+} }}
\newcommand{\dsstm}  {\ensuremath{ D_s^{*-} }}

\newcommand{\dsds}     {\ensuremath{ \dsp  \dsm    }}
\newcommand{\dsdsst}   {\ensuremath{ \dsp  \dsstm  }}
\newcommand{\dsstdsst} {\ensuremath{ \dsstp \dsstm }}

\newcommand{\dspn}  {\ensuremath{ D_s^{(*)+} }}
\newcommand{\dsmn}  {\ensuremath{ D_s^{(*)-} }}
\newcommand{\dsdsn} {\ensuremath{ \dspn \dsmn }}

\newcommand{\eedsds}    {\ensuremath{ \ee \to \dsds   }}
\newcommand{\eedsdsg}   {\ensuremath{ \ee \to \dsds \gisr  }}

\newcommand{\eedsdsst}   {\ensuremath{ \ee \to \dsdsst   }}
\newcommand{\eedsdsstg}  {\ensuremath{ \ee \to \dsdsst \gisr  }}

\newcommand{\eedsstdsst}  {\ensuremath{ \ee \to \dsstdsst   }}
\newcommand{\eedsstdsstg} {\ensuremath{ \ee \to \dsstdsst \gisr  }}

\newcommand{\eedsdsn}      {\ensuremath{ \ee \to \dsdsn   }}
\newcommand{\eedsdsng}     {\ensuremath{ \ee \to \dsdsn \gisr  }}

\newcommand{\mdspn}    {\ensuremath{ M_{\dspn} }}
\newcommand{\mdsmn}    {\ensuremath{ M_{\dsmn} }}

\newcommand{\mdsp}     {\ensuremath{ M_{\dsp}   }}
\newcommand{\mdsm}     {\ensuremath{ M_{\dsm}   }}

\newcommand{\mdsds}    {\ensuremath{ M_{\dsds}     }}

\newcommand{\mdsdsn}   {\ensuremath{ M_{\dsdsn}    }}
 
\newcommand{\mdsstp}    {\ensuremath{ M_{\dsstp}    }}
\newcommand{\mdsstm}    {\ensuremath{ M_{\dsstm}    }}
\newcommand{\mdsdsst}   {\ensuremath{ M_{\dsdsst}   }}

\newcommand{\mdsstdsst} {\ensuremath{ M_{\dsstdsst} }}

\newcommand{\dn}   {\ensuremath{ D^0 }}
\newcommand{\dpl}  {\ensuremath{ D^+ }}
\newcommand{\dstm} {\ensuremath{ D^{*-} }}

\newcommand{\dpdstm}   {\ensuremath{ D^+ D^{*-}     }}
\newcommand{\dstpdstm} {\ensuremath{ D^{*+} D^{*-}  }}
\newcommand{\dndmpp}   {\ensuremath{ D^0 D^- \pi^+  }}
\newcommand{\ddstp}    {\ensuremath{ \dn  \dstm \pi^+ }}

\newcommand{\dd}       {\ensuremath{ D   \overline D      }}
\newcommand{\ddst}     {\ensuremath{ D   \overline D{}^*  }}

\newcommand{\dstdst}   {\ensuremath{ D^* \overline D{}^* }}
\newcommand{\lala}     {\ensuremath{ \Lambda_c^+ \Lambda_c^- }}

\begin{document}

\title{\quad\\[0.5cm] Measurement of \eedsdsn\ cross sections near
  threshold using initial-state radiation}

\affiliation{Budker Institute of Nuclear Physics, Novosibirsk}
\affiliation{Faculty of Mathematics and Physics, Charles University, Prague}
\affiliation{Chiba University, Chiba}
\affiliation{University of Cincinnati, Cincinnati, Ohio 45221}
\affiliation{Justus-Liebig-Universit\"at Gie\ss{}en, Gie\ss{}en}
\affiliation{The Graduate University for Advanced Studies, Hayama}
\affiliation{Gyeongsang National University, Chinju}
\affiliation{Hanyang University, Seoul}
\affiliation{University of Hawaii, Honolulu, Hawaii 96822}
\affiliation{High Energy Accelerator Research Organization (KEK), Tsukuba}
\affiliation{Hiroshima Institute of Technology, Hiroshima}
\affiliation{University of Illinois at Urbana-Champaign, Urbana, Illinois 61801}
\affiliation{Indian Institute of Technology Guwahati, Guwahati}
\affiliation{Institute of High Energy Physics, Chinese Academy of Sciences, Beijing}
\affiliation{Institute of High Energy Physics, Vienna}
\affiliation{Institute of High Energy Physics, Protvino}
\affiliation{Institute for Theoretical and Experimental Physics, Moscow}
\affiliation{J. Stefan Institute, Ljubljana}
\affiliation{Kanagawa University, Yokohama}
\affiliation{Institut f\"ur Experimentelle Kernphysik, Karlsruher Institut f\"ur Technologie, Karlsruhe}
\affiliation{Korea Institute of Science and Technology Information, Daejeon}
\affiliation{Korea University, Seoul}
\affiliation{Kyungpook National University, Taegu}
\affiliation{\'Ecole Polytechnique F\'ed\'erale de Lausanne (EPFL), Lausanne}
\affiliation{Faculty of Mathematics and Physics, University of Ljubljana, Ljubljana}
\affiliation{University of Maribor, Maribor}
\affiliation{Max-Planck-Institut f\"ur Physik, M\"unchen}
\affiliation{University of Melbourne, School of Physics, Victoria 3010}
\affiliation{Nagoya University, Nagoya}
\affiliation{Nara Women's University, Nara}
\affiliation{National Central University, Chung-li}
\affiliation{National United University, Miao Li}
\affiliation{Department of Physics, National Taiwan University, Taipei}
\affiliation{H. Niewodniczanski Institute of Nuclear Physics, Krakow}
\affiliation{Nippon Dental University, Niigata}
\affiliation{Niigata University, Niigata}
\affiliation{University of Nova Gorica, Nova Gorica}
\affiliation{Novosibirsk State University, Novosibirsk}
\affiliation{Osaka City University, Osaka}
\affiliation{Panjab University, Chandigarh}
\affiliation{University of Science and Technology of China, Hefei}
\affiliation{Seoul National University, Seoul}
\affiliation{Sungkyunkwan University, Suwon}
\affiliation{School of Physics, University of Sydney, NSW 2006}
\affiliation{Tata Institute of Fundamental Research, Mumbai}
\affiliation{Excellence Cluster Universe, Technische Universit\"at M\"unchen, Garching}
\affiliation{Toho University, Funabashi}
\affiliation{Tohoku Gakuin University, Tagajo}
\affiliation{Tohoku University, Sendai}
\affiliation{Department of Physics, University of Tokyo, Tokyo}
\affiliation{Tokyo Metropolitan University, Tokyo}
\affiliation{Tokyo University of Agriculture and Technology, Tokyo}
\affiliation{CNP, Virginia Polytechnic Institute and State University, Blacksburg, Virginia 24061}
\affiliation{Wayne State University, Detroit, Michigan 48202}
\affiliation{Yonsei University, Seoul}
  \author{G.~Pakhlova}\affiliation{Institute for Theoretical and Experimental Physics, Moscow} 
  \author{I.~Adachi}\affiliation{High Energy Accelerator Research Organization (KEK), Tsukuba} 
  \author{H.~Aihara}\affiliation{Department of Physics, University of Tokyo, Tokyo} 
  \author{K.~Arinstein}\affiliation{Budker Institute of Nuclear Physics, Novosibirsk}\affiliation{Novosibirsk State University, Novosibirsk} 
 \author{T.~Aushev}\affiliation{\'Ecole Polytechnique F\'ed\'erale de Lausanne (EPFL), Lausanne}\affiliation{Institute for Theoretical and Experimental Physics, Moscow} 
  \author{T.~Aziz}\affiliation{Tata Institute of Fundamental Research, Mumbai} 
  \author{A.~M.~Bakich}\affiliation{School of Physics, University of Sydney, NSW 2006} 
  \author{V.~Balagura}\affiliation{Institute for Theoretical and Experimental Physics, Moscow} 
  \author{E.~Barberio}\affiliation{University of Melbourne, School of Physics, Victoria 3010} 
  \author{A.~Bay}\affiliation{\'Ecole Polytechnique F\'ed\'erale de Lausanne (EPFL), Lausanne} 
% ...

\collaboration{The Belle Collaboration}

\begin{abstract}
We report a measurement of exclusive \eedsdsn\ cross sections as a
function of center-of-mass energy near \dsdsn\ threshold with
initial-state radiation. The analysis is based on a data sample
collected with the Belle detector with an integrated luminosity of
$967\,\mathrm{fb}^{-1}$.
\end{abstract}


\maketitle
\setcounter{footnote}{0}

Recently a number of measurements of exclusive cross sections for
\ee\ annihilation into charmed hadron pairs above open-charm threshold
were performed by the B-factory experiments using initial-state
radiation (ISR). These include Belle measurements~\cite{chcon} of
\ee\ cross sections for the \dd\ ($D=\dn$ or \dpl), \dpdstm,
\dstpdstm, \dndmpp, \ddstp\ and \lala\ final
states~\cite{belle:dd,belle:ddst,belle:ddp,belle:ddstp,belle:ll} and
BaBar measurements of the \dd, \ddst, \dstdst\ final
states~\cite{babar:dd,babar:ddst}, which are, in general, consistent
with those of Belle.  In addition, CLEO scanned the \ee\ center of
mass energy range from 3.97 to 4.26\gev\ and obtained exclusive cross
sections for the \dd, \ddst, \dstdst, $\dd\pi$ and $\ddst\pi$ final
states~\cite{cleo:cs}.  The first measurements of the exclusive cross
sections for \ee\ annihilation into {\it charmed strange} meson pairs
\dsdsn\ were performed by CLEO with high accuracy but with limited
maximum energy (4.26\gev )~\cite{cleo:cs}. Recently BaBar presented
\eedsdsn\ cross sections averaged over 100\mev\ wide
bins~\cite{babar:dsds}.  The observed cross sections were found to be
an order of magnitude smaller than those for non-strange charmed meson
production.

Although the recent BES fit to the total cross section for hadron
production in \ee\ provided new parameter values for the $\psi$
resonances~\cite{bes:fit}, the available {\it exclusive} \ee\ cross
sections have not yet been qualitatively explained. One of the main
problems is the numerous open charm thresholds in the region that
influence the cross section behavior and, thus, complicate theoretical
descriptions.

The $Y$ states~\cite{y_states} (with masses above open charm threshold
and quantum numbers $J^{PC}= 1^{- -} $), which do not exhibit strong
decays to any of the measured open charm final states and have remain
unexplained since their discovery more than five years, provide
additional motivation to pursue all possible experimental information
about the decomposition of charmed particle production in the
charm-threshold region.

Here we report a measurement of exclusive \eedsdsn\ cross sections as
a function of center-of-mass energy from the \dsdsn\ thresholds to
5.0\gev, continuing our studies of the exclusive open charm production
in this mass range. The analysis is based on a study of events with
ISR photons in a data sample collected with the Belle
detector~\cite{det} at the $\Upsilon(nS)$ ($n=1,..., 5$) resonances
and nearby continuum with an integrated luminosity of
$967\,\mathrm{fb}^{-1}$ at the KEKB~\cite{kekb} asymmetric-energy
\ee\ collider.

We follow the full reconstruction method that was previously used for
the measurements of the \ee\ cross sections to \dd, \dndmpp\ and
\ddstp\ final states~\cite{belle:dd,belle:ddp,belle:ddstp}.  We select
\eedsdsng\ signal events in which the \dspn\ and \dsmn\ mesons are
fully reconstructed.  ISR photon candidates are indicated by \gisr.
In general, an \gisr\ is not required to be detected and its presence
in the event is inferred from a peak at zero in the spectrum of recoil
mass squared against the \dsdsn\ system.  The recoil mass squared is
defined as:
\begin{eqnarray}
\RMS(\dsdsn)=(\ecm - E_{\dsdsn})^2 - \nonumber \\  p^2_{\dsdsn} ,
\end{eqnarray}
where \ecm\ is the initial \ee\ center-of-mass ($\mathrm{c.m.}$)
energy, $E_{\dsdsn}$ and $p_{\dsdsn}$ are energy and three-momentum of
the \dsdsn\ combination, respectively.  To suppress backgrounds two
cases are considered: (1) the \gisr\ is outside of the detector
acceptance and the polar angle for the \dsdsn\ combination in the
$\mathrm{c.m.}$ frame is in the range $|\cos(\theta_{\dsdsn})|>0.9$;
(2) the \gisr\ is within the detector acceptance
($|\cos(\theta_{\dsdsn})|<0.9$). In the latter case, the \gisr\ is
required to be detected and the mass of the $\dsdsn\gisr$ combination
should to be greater than ($\ecm -0.5\gev$).  To suppress backgrounds
from $\ee \to \dsdsn (m)(\pi^+\pi^-)\gisr, (m=1,2,...)$ processes, we
exclude events that contain additional charged tracks that were not
used in the \dspn\ and the \dsmn\ reconstruction.

All charged tracks are required to originate from the vicinity of the
interaction point (IP); we impose the requirements $|dr|<1 \,
{\mathrm{cm}}$ and $|dz|<4\,{\mathrm{cm}}$, where $dr$ and $dz$ are
the impact parameters perpendicular to and along the beam direction
with respect to the IP, respectively. Charged kaons are required to
have a ratio of particle identification likelihood, $\mathcal{P}_K =
\mathcal{L}_K / (\mathcal{L}_K + \mathcal{L}_\pi)>0.6$~\cite{nim}.  No
identification requirements are applied for pion candidates.

$K^0_S$ candidates are reconstructed from $\pi^+ \pi^-$ pairs with an
invariant mass within $10\mevc$ of the $K^0_S$ mass. The distance
between the two pion tracks at the $K^0_S$ vertex must be less than
$1\,\mathrm{cm}$, the transverse flight distance from the interaction
point is required to be greater than $0.1\,\mathrm{cm}$, and the angle
between the $K^0_S$ momentum direction and the flight direction in the
$x-y$ plane should be smaller than $0.1\,\mathrm{rad}$.

Photons are reconstructed in the electromagnetic calorimeter as
showers with energies greater than $50 \mev$ that are not associated
with charged tracks. Pairs of photons are combined to form $\pi^0$
candidates. If the mass of a $\gamma \gamma$ pair lies within
$15\mevc$ of the $\pi^0$ mass, the pair is fitted with a $\pi^0$ mass
constraint and considered as a $\pi^0$ candidate.  ISR photon
candidates are required to have energies greater than $2.5 \gev$.
Photon candidates used in $\eta$, $\eta^{\prime}$ and
\dsstp\ reconstruction are required to have energies greater than $100
\mev$.

$\eta$ candidates are reconstructed using $\pi^+ \pi^- \pi^0$ ($\pm
10\mevc$ mass window) and $\gamma \gamma$ ($\pm 20\mevc$ mass window)
decay modes ($\sim 2.5\,\sigma$ in each case). $\eta^{\prime}$
candidates are reconstructed using $\eta \pi^+ \pi^-$ ($\pm 10\mevc$
mass window) and $\gamma \pi^+ \pi^- $ ($\pm 15\mevc$ mass window)
decay modes ($\sim 2.0\,\sigma$ in each case). A mass- and
vertex-constrained fit is applied to $\eta$ and $\eta^{\prime}$
candidates.

\dsp\ candidates are reconstructed using six decay modes: $K^0_S K^+$,
$K^- K^+ \pi^+$, $K^- K^+ \pi^+ \pi^0$, $K^0_S K^- \pi^+ \pi^+$, $\eta
\pi^+$ and $\eta^{\prime}\pi^+$. Before calculation of the
\dsp\ candidate mass, a vertex fit to a common vertex is performed for
tracks that form the \dsp\ candidate. A $\pm 15\mevc$ mass signal
window is used for all modes ($\sim 3\,\sigma$ in each case).  To
improve the momentum resolution of \dsp\ meson candidates, the tracks
from the \dsp\ candidate are fitted to a common vertex with a mass
\dsp\ mass constraint. \dsstp\ candidates are reconstructed using the
$\dsp\gamma$ decay mode. A $\pm 15\mevc$ mass window is used ($\sim
2.5\,\sigma$). A mass-constrained fit is also applied to
\dsstp\ candidates.

The \dsp\ and \dsstp\ sidebands used for background studies are four
times as large as the signal region and are divided into windows of
the same width as that of the signal. To avoid signal
over-subtraction, the selected sidebands are shifted by $30\mevc$ from
the signal region.  The $D_s(D_s^*)$ candidates from these sidebands
are refitted to the central mass value of each window.

\begin{figure}[htb]
\hspace*{-0.025\textwidth}
\includegraphics[width=0.49\textwidth]{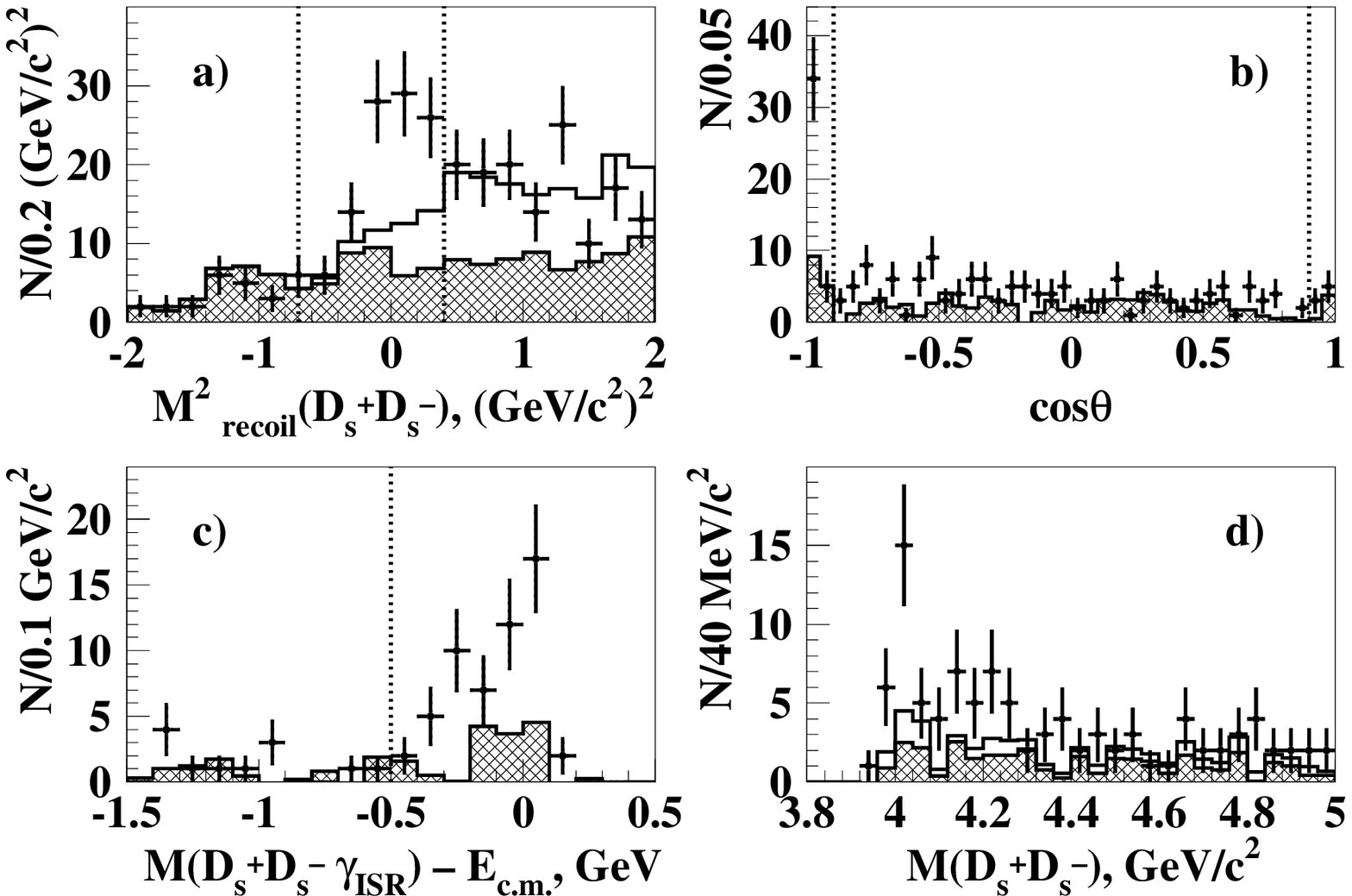} 
\caption{a) The distribution of $\RMS(\dsds)$ for $\mdsds<5.0\gevc$
  after all the requirements are applied. b) The polar angle
  distribution of the \dsds\ combinations.  c) The mass spectrum of
  the $\dsds\gisr$ combinations after subtraction of \ecm\ energy in
  case (2). d) The \mdsds\ spectrum after all the requirements
  applied. Cross-hached histograms show the normalized \mdsp\ and
  \mdsm\ sideband contributions.  Feed down from the \dsdsst\ final
  state is shown by the open histograms.  The signal windows are shown
  by vertical dashed lines.}
\label{rmx_dsds}
\end{figure}

The $\RMS(\dsds)$ distribution for $\mdsds<5.0\gevc$ after all the
requirements described above is shown in Fig.~\ref{rmx_dsds}\,a). A
clear peak corresponding to the \eedsdsg\ process is evident near zero
recoil mass. The shoulder at positive values is mainly due to $\ee \to
\dsdsst\gisr$ background. We define the signal region with an
asymmetric requirement $-0.7(\gevc)^2<\RMS(\dsds)<0.4(\gevc)^2$ to
suppress event this background. The polar angle distribution of the
\dsds\ combinations and the mass spectrum of the $\dsds\gisr$
combinations (after subtraction of \ecm) in case (2), after
$\RMS(\dsds)$ requirement, are shown in Figs.~\ref{rmx_dsds}\,b,\,c).
These distributions are in agreement with a MC simulation and are
typical of ISR production.  The \mdsds\ spectrum with all the
requirements applied is shown in Fig.~\ref{rmx_dsds}\,d). A clear peak
is seen at threshold near the $\psi(4040)$ mass.

\begin{figure}[htb]
\hspace*{-0.025\textwidth}
\includegraphics[width=0.49\textwidth]{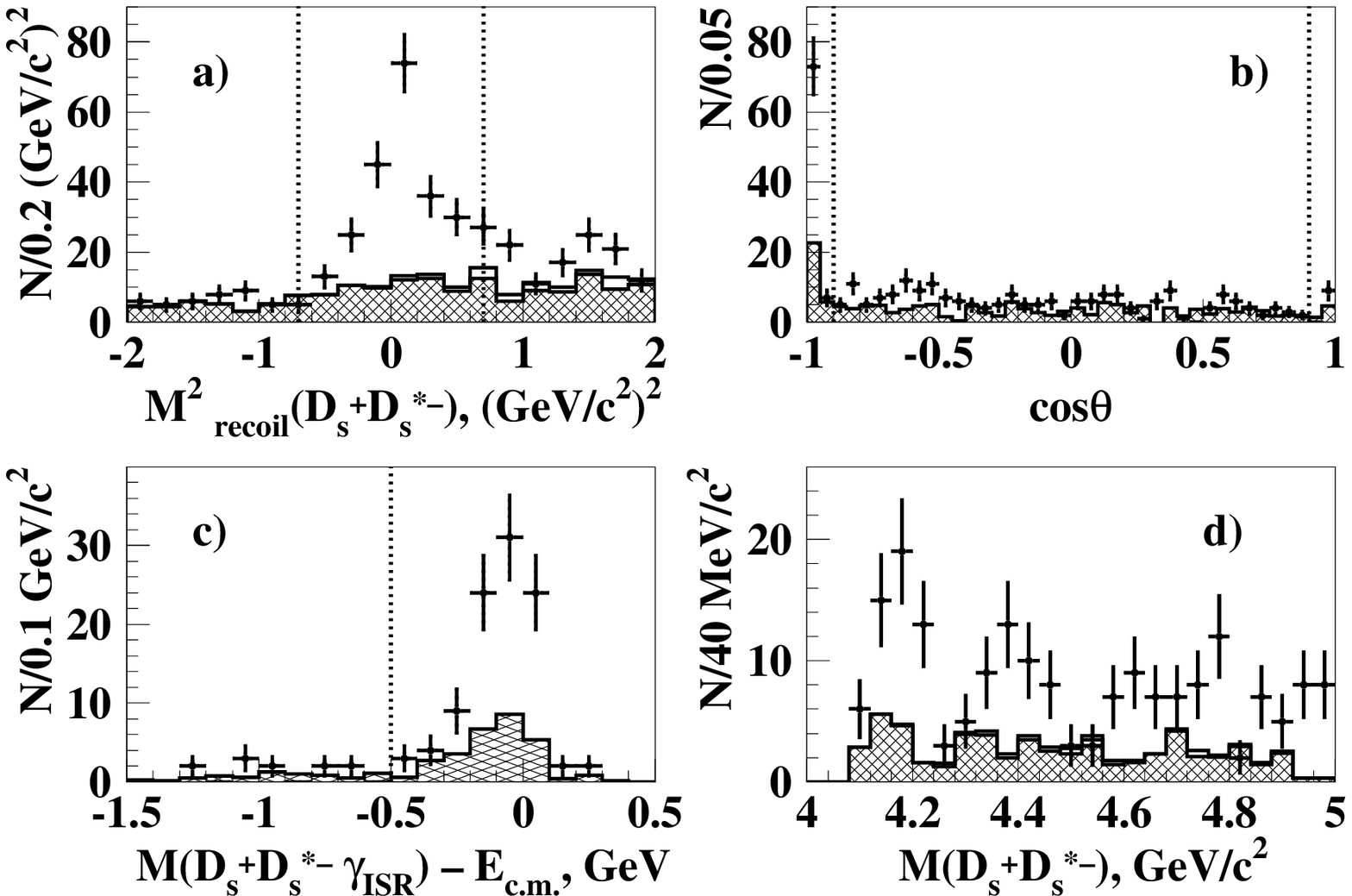} 
\caption{a) The distribution of the $\RMS(\dsdsst)$ for
  $\mdsdsst<5.0\gevc$ after all the requirements applied. b) The polar
  angle distribution of the \dsdsst\ combinations. c) The mass
  spectrum of the $\dsdsst\gisr$ combinations after subtraction of
  \ecm\ in case (2). d) The obtained \mdsdsst\ spectrum.  Cross-hached
  histograms show the normalized \mdsp\ and \mdsstm\ sidebands
  contributions. The small contamination from the \dsstdsst\ final
  state is shown by the open histograms. The signal windows are shown
  by vertical dashed lines.}
\label{rmx_dsdst}
\end{figure}

The $\RMS(\dsdsst)$ distribution for $\mdsdsst<5.0\gevc$ after all the
requirements are applied is shown in Fig.~\ref{rmx_dsdst}\,a). A clear
peak corresponding to \eedsdsstg\ signal process is again evident
around zero. We define the signal region for $\RMS(\dsdsst)$ by a
requirement $\pm 0.7(\gevc)^2$ around zero. The polar angle
distribution of the \dsdsst\ combinations and the mass spectrum of
$\dsdsst\gisr$ combinations (after subtraction of the \ecm\ energy) in
case (2) after the requirement on $\RMS(\dsdsst)$ is applied are shown
in Figs.~\ref{rmx_dsdst}\,b,\,c).  The \mdsdsst\ spectrum after all
the requirements are applied is shown in Fig.~\ref{rmx_dsdst}\,d). Two
clear peaks are seen at the $\psi(4160)$ and the $\psi(4415)$ masses.

\begin{figure}[htb]
\hspace*{-0.025\textwidth}
\includegraphics[width=0.49\textwidth]{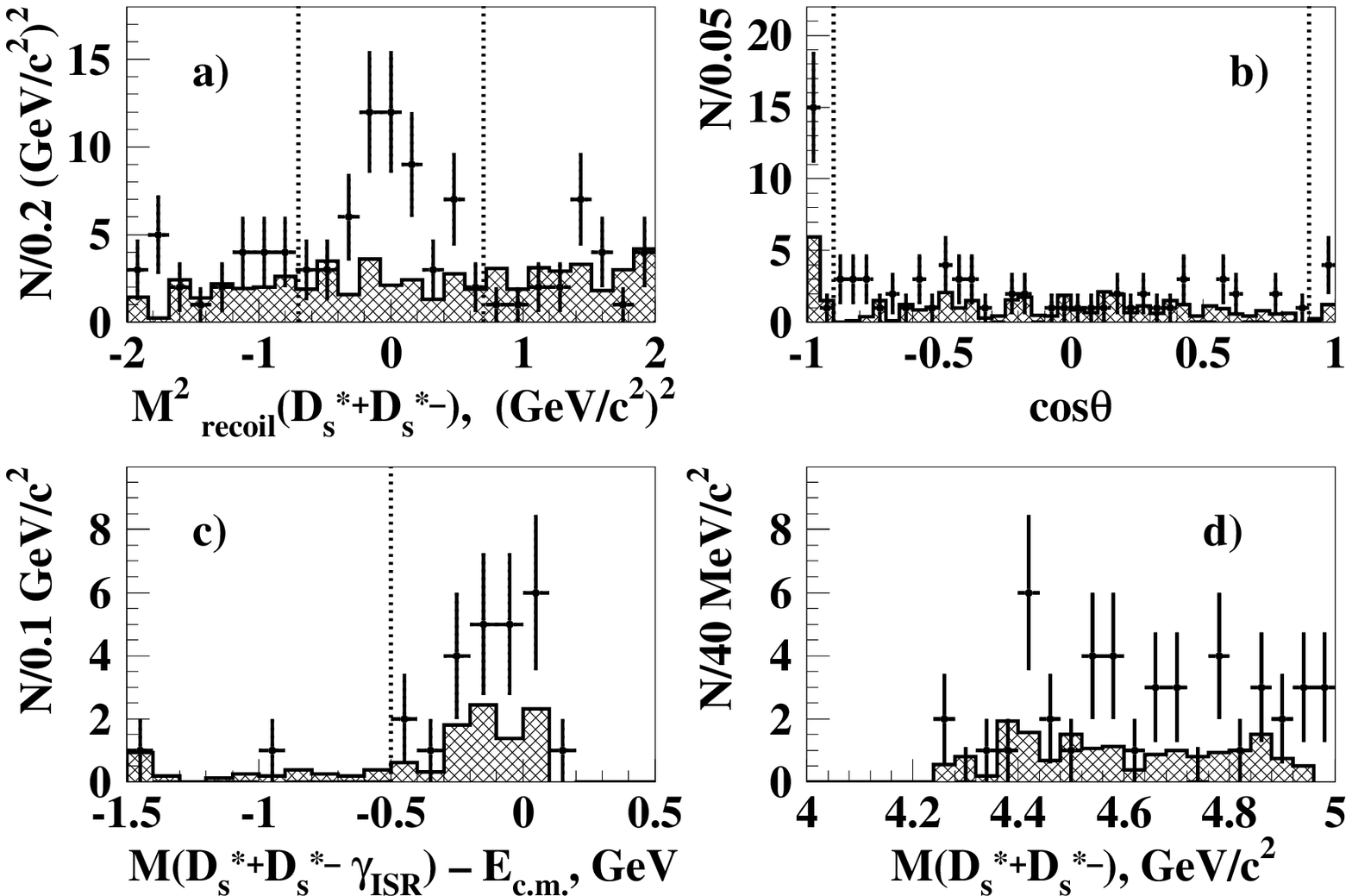} 
\caption{a) The distributions of the $\RMS(\dsstdsst)$ for
  $\mdsstdsst<5.0\gevc$ after all the requirements applied.  b) The
  polar angle distribution of the \dsstdsst\ combinations. c) The mass
  spectrum of the $\dsstdsst\gisr$ combinations after subtraction of
  \ecm\ in case (2).  d) The \mdsstdsst\ spectrum after all the
  requirements applied. Cross-hached histograms show the normalized
  \mdsstp\ and \mdsstm\ sidebands contributions.  The signal windows
  are shown by vertical dashed lines.}
\label{rmx_dstdst}
\end{figure}

The $\RMS(\dsstdsst)$ distribution for $\mdsstdsst<5.0\gevc$ after all
the requirements applied is shown in Fig.~\ref{rmx_dstdst}\,a). A peak
corresponding to \eedsstdsstg\ is again evident around zero recoil
mass. We define the signal region for $\RMS(\dsstdsst)$ by a
requirement $\pm 0.7(\gevc)^2$ around zero recoil mass.  The polar
angle distribution of the \dsstdsst\ combinations and the mass
spectrum of the $\dsstdsst\gisr$ combinations (after subtraction of
\ecm) that survive the $\RMS(\dsstdsst)$ requirement in case (2) are
shown in Figs.~\ref{rmx_dstdst}\,b,\,c).  The full
\mdsstdsst\ spectrum after all the requirements are applied is shown
in Fig.~\ref{rmx_dstdst}\,d). With such limited statistics no
structures are evident.

The contribution of multiple entries after all the requirements is
found to be $\sim 6\%$, $\sim 22\%$, $\sim 23\%$, for the \dsds,
\dsdsst\ and the \dsstdsst\ final states, respectively. In such cases,
only one \dsdsn\ combination per event, that with the minimum value of
$\chi^2_{\mathrm{tot}} = \chi^2_{M(\dsp)} + \chi^2_{M(\dsm)} +
(\chi^2_{M(\dsstp)}) + (\chi^2_{M(\dsstm)}$), is used, where
$\chi^2_{M(\dsp)}$, $\chi^2_{M(\dsm)}$, $\chi^2_{M(\dsstp)}$ and
$\chi^2_{M(\dsstm)}$ correspond to the mass fits for the \dsp, \dsm,
\dsstp\ and the \dsstm\ candidates.

The following sources of background are considered: (1) combinatorial
background under the \dspn(\dsmn) peak combined with a correctly
reconstructed \dsmn(\dspn) from the signal or other processes; (2)
both the \dspn\ and the \dsmn\ are combinatorial; (3) for the
\dsds\ final state: reflections from the \eedsdsstg\ and
\eedsstdsstg\ processes followed by $D_s^* \to D_s \gamma$, where the
low momentum $\gamma$ is not reconstructed; for the \dsdsst\ final
state: reflection from the \eedsstdsstg\ process followed by $D_s^*
\to D_s \gamma$, where the low momentum $\gamma$ is not reconstructed;
(4) reflection from the $\eedsdsn\pi^0_{miss}\gisr$ processes, where
the $\pi^0_{miss}$ is not reconstructed.  (5) the contribution of
$\eedsdsn\pi^0$, when an energetic $\pi^0$ is misidentified as a
single \gisr.

The contribution from background (1) is extracted using the \dsmn\ and
\dspn\ sidebands.  Background (2) is present in both the \mdspn\ and
\mdsmn\ sidebands and is, thus, subtracted twice. To account for this
over-subtraction we use a double sideband region, when events are
selected from both the \mdspn\ and the \mdsmn\ sidebands.  The
contributions from the combinatorial backgrounds (1--2) are shown in
Figs.~\ref{rmx_dsds}\,d) and~\ref{rmx_dsdst}\,d) as cross-hatched
histograms.

Backgrounds (3--4) are suppressed by the tight requirement on
$\RMS(\dsdsn)$.  The remaining background (3) for the \dsdsst\ final
state is estimated using a MC simulation of the \eedsstdsstg\ process.
To reproduce the shape of the \dsdsst\ mass spectrum we use the
initial measurement of the \dsstdsst\ mass spectrum.  The remainder of
background (3) for the \dsds\ final state is estimated using a MC
simulation of the \eedsdsstg\ and \eedsstdsstg\ processes.  To
reproduce the shape of the \dsds\ mass spectrum we use the initial
measurement of the \dsstdsst\ mass spectrum and the first iteration of
the \dsdsst\ mass spectrum.  The contributions from background (3) for
the \dsds\ and the \dsdsst\ final states are shown in
Figs.~\ref{rmx_dsds}\,a),~\ref{rmx_dsds}\,d) and
Figs.~\ref{rmx_dsdst}\,a),~\ref{rmx_dsdst}\,d) as open
histograms. Uncertainties in these estimates are included in the
systematic errors.

To estimate the contribution from background (4), we study the
$\eedsdsn\pi^0\gisr$ processes using fully reconstructed final states.
From a MC study we estimate the fraction of reconstructed events for
the cases where the $\pi^0$ is not detected. After the application of
the requirement on $\RMS(\dsdsn)$ this contribution is found to be
less than 0.5\% and negligibly small; uncertainties in this estimate
are included in the systematic errors.

The contribution from background (5), in which an energetic $\pi^0$ is
misidentified as the \gisr\ candidate, is determined from the data
using fully reconstructed $\eedsdsn\pi^0$ events.  Only three events
with $\mdsds<5.0\gevc$ and $M_{\dsds\pi^0}-\ecm>0.5\gev$ are found in
the data. Assuming a uniform $\pi^0$ polar angle distribution, this
background contribution in the $|\cos(\theta_{\dsds})|>0.9$ signal
sub-sample (case 1) is 3\,events/9$\epsilon_{\pi^0} \sim 0.6\,$ events
in the whole \mdsds\ mass range, where $\epsilon_{\pi^0}$ is the
$\pi^0$ reconstruction efficiency. For the \dsdsst\ and the
\dsstdsst\ final states the expected backgrounds are $\sim 0.6\,$
events and 0 events in the whole \mdsdsst\ and \mdsstdsst\ mass
ranges.  The probability of $\pi^0 \to \gamma$ misidentification due
to asymmetric $\pi^0 \to \gamma \gamma$ decays is also estimated to be
small. Thus the contribution from background (5) is found to be
negligibly small; uncertainties in these estimates are included in the
systematic error.

The \eedsdsn\ cross sections are extracted from the background
subtracted \dsdsn\ mass distributions
\begin{eqnarray}
\sigma(\eedsdsn) = \frac{ dN/dm }{ \eta_{\mathrm{tot}} dL/dm} \, ,
\end{eqnarray}
where $m\equiv\mdsdsn$, $dN/dm$ is the obtained mass spectrum,
$\eta_{\mathrm{tot}}$ is the total efficiency~\cite{cs}.  The factor
$dL/dm$ is the differential ISR luminosity:
\begin{eqnarray}
 dL/dm =\frac{\alpha}{\pi x}\Bigl((2-2x+x^2)\ln\frac{1+C}{1-C}-x^2C\Bigr)
\frac{2m \mathcal{L}}{{\ecm}^2} \, ,
\end{eqnarray}
where $x = 1 - m^2/{\ecm}^2$, $\mathcal{L}$ is the total integrated
luminosity and $C = \cos\theta_0$, where $\theta_0$ denotes the polar
angle range for \gisr\ in the \ee\ c.m. frame:
$\theta_0<\theta_{\gisr}<180-\theta_0$.  The total efficiencies
determined by the MC simulation grow quadratically with energy from
0.015\%, 0.010\%, 0.005\% near threshold to 0.045\%, 0.025\%, 0.011\%
at 5.0\gevc\ for the \dsds, \dsdsst\ and the \dsstdsst\ final states,
respectively.  The resulting \eedsdsn\ exclusive cross sections
averaged over the bin width are shown in Fig.~\ref{cs}. Since the bin
width is much larger than the \mdsdsn\ resolution, which varies from
$\sim2\mevc$ around threshold to $\sim6\mevc$ at $\mdsdsn=5.0\gevc$,
no correction for resolution is applied.  The next-to-leading order
radiative corrections are taken into account by the $dL/dm$
formula. The next-to-next-to-leading order corrections are included in
the systematics. The contribution of final state radiation (FSR) is
strongly suppressed~\cite{quest} and is neglected in this study.
\onecolumngrid \appendix
\begin{figure}[htb]
\hspace*{-0.025\textwidth}
\includegraphics[width=0.99\textwidth]{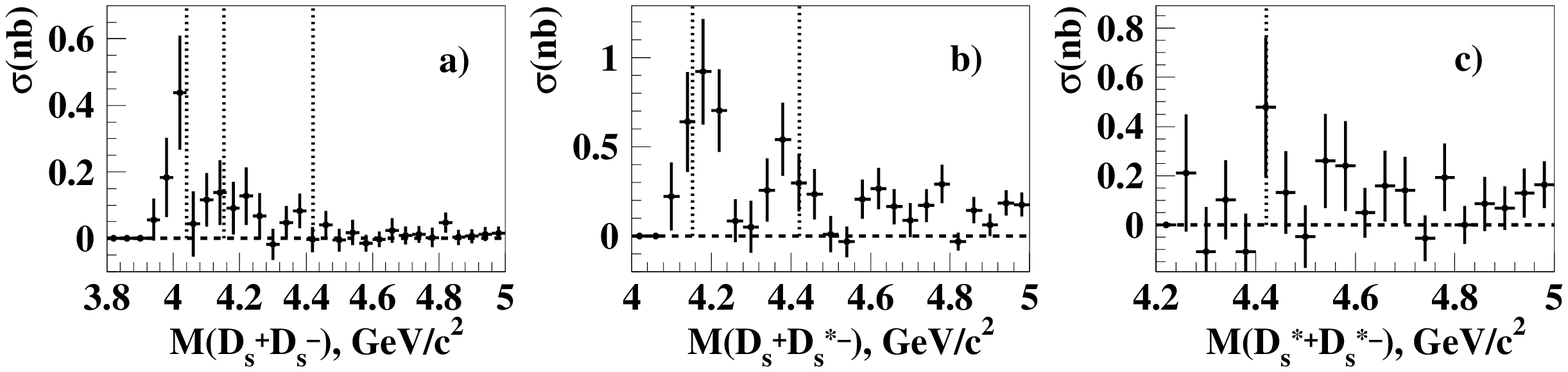} 
\caption{The cross section averaged over the bin width for a) the
  \eedsds\ process; b) the \eedsdsst\ + c.c. process; c) the
  \eedsstdsst\ process. Error bars show statistical uncertainties
  only. There is a common systematic uncertainty for all measurements,
  11\% for \dsds, 17\% for \dsdsst\ and 31\% for \dsstdsst.  This
  uncertainty is described in the text. The dotted lines show masses of
  the $\psi(4040)$, $\psi(4160)$ and $\psi(4415)$ states~\cite{pdg}.}
\label{cs}
\end{figure}
\twocolumngrid

\begin{figure}[htb]
\begin{tabular}{c}
\hspace*{-0.025\textwidth}
\includegraphics[width=0.49\textwidth]{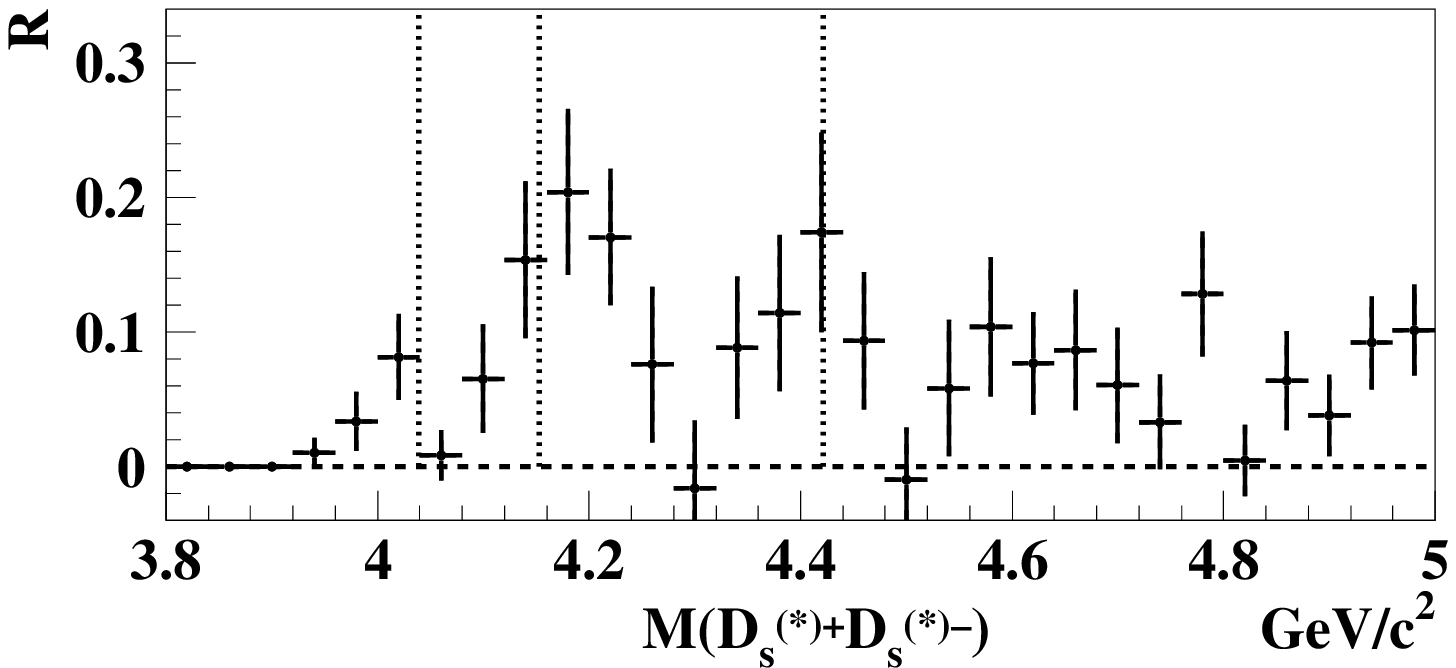} 
\end{tabular}
\caption{The R ratio for the sum of the \dsds, \dsdsst\ and the
  \dsstdsst\ final states. The vertical dotted lines show the masses
  of the $\psi(4040)$, $\psi(4160)$ and $\psi(4415)$
  states~\cite{pdg}. Error bars show statistical uncertainties only.}
\label{r_sum}
\end{figure}

 The R ratio, defined as $R=\sigma(\ee\to hadrons)/
\sigma(\ee\to\mu^+\mu^-)$, where $\sigma(\ee\to\mu^+\mu^-) =
4\pi\alpha^2/3s$, for the sum of the exclusive \eedsdsn\ cross
sections is shown in Fig.~\ref{r_sum}.

The systematic errors for the $\sigma(\eedsdsn)$ measurements are
summarized in Table ~\ref{tab1}.
\begin{table}[htb]
\caption{Contributions to the systematic error on the cross sections.}
\label{tab1}
\begin{center}
\begin{tabular}{@{\hspace{0.2cm}}l@{\hspace{0.2cm}}@{\hspace{0.2cm}}c@{\hspace{0.2cm}}@{\hspace{0.2cm}}c@{\hspace{0.2cm}}@{\hspace{0.2cm}}c@{\hspace{0.2cm}}}
\hline\hline
Source & \dsds & \dsdsst & \dsstdsst \\
\hline
Background subtraction    & $\pm 4\%$ & $\pm 7\%$  & $\pm 24\%$ \\
Cross section calculation & $\pm 7\%$ & $\pm 11\%$ & $\pm 12\%$ \\
$\mathcal{B}(D_s^{(*)})$   & $\pm 5\%$ & $\pm 5\%$  & $\pm 5\%$  \\
Reconstruction            & $\pm 6\%$ & $\pm 10\%$  & $\pm 16\%$  \\
Kaon identification       & $\pm 2\%$ & $\pm 2\%$  & $\pm 2\%$  \\
\hline
Total                     & $\pm 11\%$ & $\pm 17\%$ & $\pm 31\%$ \\ 
\hline\hline
\end{tabular}
\end{center}
\end{table}
The systematic errors associated with the background (1--2)
subtraction are estimated from the uncertainty in the scaling factors
for the sideband subtractions. This is done using fits to the
\mdspn\ and \mdsmn\ distributions in the data with different signal
and background parameterizations and are found to be 3\%, 7\% and 24\%
for the \dsds, \dsdsst\ and the \dsstdsst\ final states, respectively.
Uncertainties in the contribution from background (3) are estimated to
be 2\% for the \dsds\ final state and smaller than 1\% for the
\dsdsst\ final state.  Uncertainties in the backgrounds (4--5) are
estimated conservatively to be smaller than 1\% of the signal for all
final states. The systematic errors ascribed to the cross section
calculation include an error on the differential ISR luminosity (2\%)
and statistical errors of the MC simulation for the total efficiency
calculations. In the case of the \dsdsst\ state there is an additional
uncertainty due to the unknown helicity distribution, estimated
following a procedure similar to that used in
Ref~\cite{belle:ddst}. Another source of systematic error comes from
the uncertainties in track and photon reconstruction efficiencies (1\%
per track, 1.5\% per photon and 7\% per soft photon). Other
contributions include the uncertainty in the identification efficiency
and the absolute \dsp\ and \dsstp\ branching fractions~\cite{pdg}. The
total systematic uncertainties are 11\%, 17\% and 31\% for the \dsds,
\dsdsst\ and the \dsstdsst\ final states, respectively.

In summary, we report the measurement of the \eedsdsn\ exclusive cross
sections over the center-of-mass energy range from the
\dsdsn\ thresholds to 5.0\gev. A clear peak at threshold around the
$\psi(4040)$ mass is seen in the \eedsds\ cross section. In the
\eedsdsst\ cross section two peaks are evident around the $\psi(4160)$
and the $\psi(4415)$ masses.  The limited statistics do not reveal any
structures in the \eedsstdsst\ cross section. The obtained R ratio for
the sum of \eedsdsn\ cross sections has a rich structure including
peaks around the $\psi(4040)$, $\psi(4160)$ and the $\psi(4415)$
masses. Both the \eedsdsst\ cross section and the R ratio exhibit an
obvious dip near the $Y(4260)$ mass, similar to what is seen in
$\ee\to\dstdst$ and in the total cross section for charm
production. The obtained cross sections are consistent within errors
with those from BaBar~\cite{babar:dsds}. The CLEO exclusive cross
sections~\cite{cleo:cs} are not radiatively corrected and, therefore,
cannot be directly compared to the results reported here.

In this study we do not perform a fit to the obtained \eedsdsn\ cross
sections. The numerous open charm thresholds in this region complicate
the cross sections behavior and coupled-channel modifications to the
description of any particular final state require one to take into
account all other final states contributing to the total cross section
for charm production.

We thank the KEKB group for the excellent operation of the
accelerator, the KEK cryogenics group for the efficient operation of
the solenoid, and the KEK computer group and the National Institute of
Informatics for valuable computing and SINET3 network support.  We
acknowledge support from the Ministry of Education, Culture, Sports,
Science, and Technology (MEXT) of Japan, the Japan Society for the
Promotion of Science (JSPS), and the Tau-Lepton Physics Research
Center of Nagoya University; the Australian Research Council and the
Australian Department of Industry, Innovation, Science and Research;
the National Natural Science Foundation of China under contract
No.~10575109, 10775142, 10875115 and 10825524; the Ministry of
Education, Youth and Sports of the Czech Republic under contract
No.~LA10033 and MSM0021620859; the Department of Science and
Technology of India; the BK21 and WCU program of the Ministry
Education Science and Technology, National Research Foundation of
Korea, and NSDC of the Korea Institute of Science and Technology
Information; the Polish Ministry of Science and Higher Education; the
Ministry of Education and Science of the Russian Federation and the
Russian Federal Agency for Atomic Energy; the Slovenian Research
Agency; the Swiss National Science Foundation; the National Science
Council and the Ministry of Education of Taiwan; and the
U.S.\ Department of Energy.  This work is supported by a Grant-in-Aid
from MEXT for Science Research in a Priority Area (``New Development
of Flavor Physics''), and from JSPS for Creative Scientific Research
(``Evolution of Tau-lepton Physics'').


\begin{thebibliography} {99}

\bibitem{chcon} Charge-conjugate modes are included throughout this
  paper.

\bibitem{belle:dd} G.~Pakhlova {\it {et al.}} (Belle Collab.),
  Phys. Rev. D {\bf 77}, 011103 (2008).

\bibitem{belle:ddst} G.~Pakhlova {\it et al.} (Belle Collab.),
  Phys. Rev. Lett. {\bf 98}, 092001 (2007).

\bibitem{belle:ddp} G.~Pakhlova {\it et al.} (Belle Collab.),
  Phys. Rev. Lett. {\bf 100}, 062001 (2008).

\bibitem{belle:ddstp} G.~Pakhlova {\it et al.} (Belle Collab.),
  Phys. Rev. D {\bf 80}, 091101 (2009).

\bibitem{belle:ll} G.~Pakhlova {\it et al.} (Belle Collab.),
   Phys. Rev. Lett. {\bf 101}, 172001 (2008).

\bibitem{babar:dd} B.~Aubert {\it et al.} (BaBar Collab.),
  Phys. Rev. D {\bf 76}, 111105 (2007).

\bibitem{babar:ddst} B.~Aubert {\it et al.} (BaBar Collab.),
  Phys. Rev. D {\bf 79}, 092001 (2009).

\bibitem{cleo:cs} D.~Cronin-Hennessy {\it et al.} (CLEO
  Collab.),   Phys. Rev. D {\bf 80}, 072001 (2009).

\bibitem{babar:dsds} P.~del~Amo~Sanchez {\it et al.} (BaBar Collab.),
  Phys. Rev. D {\bf 82}, 052004 (2010).

\bibitem{bes:fit} M.~Ablikim {\it et al.} (BES Collab.),
  Phys. Lett. B {\bf 660}, 315 (2008).

\bibitem{y_states} B.~Aubert {\it et al.} (BaBar Collab.),
  Phys. Rev. Lett. {\bf 95}, 142001 (2005); T.~E.~Coan {\it et al.}
  (CLEO Collab.), Phys. Rev. Lett. {\bf 96}, 162003 (2006); Q.~He {\it
    et al.} (CLEO Collab.), Phys. Rev. D {\bf 74}, 091104 (2006);
  C.~Z.~Yuan {\it et al.} (Belle Collab.), Phys. Rev. Lett. {\bf 99},
  182004 (2007); B.~Aubert {\it et al.}  (BaBar Collab.),
  arXiv:0808.1543 [hep-ex], (2008); B.~Aubert {\it et al.} (BaBar
  Collab.), Phys. Rev. Lett. {\bf 98}, 212001 (2007); X.~L.~Wang {\it
    et al.} (Belle Collab.), Phys. Rev. Lett. {\bf 99}, 142002 (2007);
  Z.~Q.~Liu, X.~S.~Qin and C.~Z.~Yuan, Phys. Rev. D {\bf 78}, 014032
  (2008).

\bibitem{det} A.~Abashian {\it et al.} (Belle Collaboration),
  Nucl. Instr. and Meth.  A {\bf 479}, 117 (2002).

\bibitem{kekb} S.~Kurokawa and E.~Kikutani, Nucl. Instr. and Meth. A
  {\bf 499}, 1 (2003); and other papers included in this volume.

\bibitem{nim} E.~Nakano, Nucl. Instrum. Meth., A {\bf 494}, 402
  (2002).

\bibitem{pdg} K.~Nakamura {\it et al.} (Particle Data Group),
  J. Phys. G {\bf 37}, 075021 (2010).

\bibitem{cs} E.~A.~Kuraev and V.~S.~Fadin, Sov. J. Nucl. Phys.  {\bf 41},
  466 (1985) [Yad. Fiz {\bf 41}, 733 (1985)].

\bibitem{quest} S.~Actis {\it et al.}, Eur. Phys. J. C {\bf 66}, 585
  (2010).

\end{thebibliography}
\end{document}